\documentclass[12pt,twoside,dvips]{article}
\usepackage{amsmath,amssymb,amsthm,latexsym}
\usepackage{layout}
\usepackage{graphicx}
\newcommand{\VH}[1]{\hat{\boldsymbol{#1}}}
\newcommand{\V}[1]{\boldsymbol{#1}}

\begin{document}
\title{\bf Bootstrap Confidence Regions for Optimal Operating Conditions
           in Response Surface Methodology}

\vspace{0.15in}

\author{
{R\sc{oger} D. \sc{Gibb}}\footnote{gibb.rd@pg.com, The Procter \& Gamble Company, Mason, Ohio 45040-9452.},\
{I-\sc{Li} L\sc{u}}\footnote{Correspondence: I-Li.Lu@boeing.com, Applied Statistics, Phantom Works, The Boeing Company, P.O. Box 3707, MC 7L-22, Seattle, WA 98124-2207.}, and
{W\sc{alter} H. C\sc{arter}, J\sc{r}}\footnote{whcarter@vcu.edu, Department of Biostatistics, Virginia Commonwealth University, Richmond, VA 23298-0032.}
}
\date{}
\maketitle
\vspace{-0.25in}
\begin{abstract}

This article concerns the application of bootstrap methodology to construct a likelihood-based confidence region for operating conditions associated with the maximum of a response surface constrained to a specified region. Unlike classical methods based on the stationary point, proper interpretation of this confidence region does not depend on unknown model parameters.  In addition, the methodology does not require the assumption of normally distributed errors. The approach is demonstrated for concave-down and saddle system cases in two dimensions. Simulation studies were performed to assess the coverage probability of these regions.

\vskip 12pt
AMS 2000 subj classification: 62F25, 62F40, 62F30, 62J05.

{\it Key words}: Stationary point; Kernel density estimator; Boundary kernel.
\end{abstract}

\baselineskip=24pt
\section{Introduction}
One of the reasons for using a response surface analysis is to determine the operating conditions
which, without loss of generality, maximize a response $y$. Development of confidence regions for
these operating conditions
has been considered by several investigators. The purpose of this report is to demonstrate that
likelihood-based bootstrap confidence region methodology is a powerful alternative which does not suffer
from some of the limitations associated with existing approaches.

Over the experimental region it is assumed that the relationship of $y$ and the regressors
$x_1,~x_2,\dots,~x_k$ can be expressed
\begin{equation} \label{GeneralModel}
y = g(\V{x},\V{\theta}) + \varepsilon ,
\end{equation}
where $g$ is an unknown continuous, differentiable function and $\varepsilon$ is a random
source of variability not accounted for in $g$. Often $g(\V{x},\V{\theta})$ can be adequately
approximated with a second-order polynomial, i.e.
\begin{equation} \label{SecondOrderModel}
  g(\V{x}) \approx \beta_0 + \V{x} \V{\beta}  +  \V{x}' \V{B} \V{x},
\end{equation}
where $\V{\beta}=(\beta_1,\beta_2,\dots,\beta_k)$ and
\begin{equation} \label{CapB}
  \V{B} =
  \begin{pmatrix}
     \beta_{11}   & \beta_{12}/2 & \dots  & \beta_{1k}/2 \\
     \beta_{12}/2 & \beta_{22}   & \dots  & \beta_{2k}/2 \\
     \vdots             &                    & \ddots & \vdots       \\
     \beta_{1k}/2 & \beta_{2k}/2 & \dots  & \beta_{kk}/2
  \end{pmatrix} .
\end{equation}

Box and Hunter (1954) constructed a confidence region for the stationary point of a second-order
response surface. The stationary point is given by
\begin{equation} \label{StatPoint}
 \V{x}_{sp} = - \frac{\V{B}^{-1} \V{\beta}}{2} ,
\end{equation}
the solution to the system of equations $\partial y/ \partial \V{x}=\V{0}$. The interpretation and
relevance of the stationary point depends on the nature of the
response surface which is determined by $\V{B}$. Consider the following cases:
1) the eigenvalues of $\V{B}$ are mixed in sign, 2) the eigenvalues of $\V{B}$ are all positive and
3) the eigenvalues of $\V{B}$ are all negative.

In cases 1 and 2 the stationary point is a saddle point and the location of minimum response,
respectively, not the location of maximum response and, therefore, is not of interest for the purpose
of this report. In case 3 the stationary point is the location of maximum response. In practice, the
model parameters, including the elements of $\V{B}$, are unknown but can be estimated from the data.
Since there is uncertainty associated with any estimator of $\V{B}$ there is also uncertainty in assessing the nature of the true response surface and, therefore, the interpretation of Box and Hunter's confidence region.

Another issue of practical importance is the data are observed over a treatment space of finite dimensions,
i.e.~the experimental region, and one is generally unwilling to extrapolate to areas
outside this region. Peterson (1992) utilized a transformation technique to develop
a confidence region for the stationary point constrained to a specified region. However, as the
approach is founded on the stationary point, the methodology suffers from the same interpretation
difficulty as Box and Hunter's confidence region.

In cases 1 and 2 the maximum response is undefined unless interest is restricted to a subset of the
treatment space. Ridge analysis was developed by Hoerl (1959) and refined by Draper (1963) to estimate
the stationary point subject to the constraint $\V{x}' \V{x} = r^2$. Under their approach the investigator
must specify a value of the Lagrangian multiplier $\mu$ which, if chosen greater than the largest
eigenvalue of $\VH{B}$, facilitates determination of the location of the predicted constrained
maximum. Stablein, Carter and Wampler (1983) constructed a confidence region for the constrained stationary
point conditional on the investigator's choice of $\mu$. However, proper interpretation of their
confidence region depends on whether the choice of $\mu$ is greater than the
largest eigenvalue of $\V{B}$. Since the eigenvalues of $\V{B}$ can only be estimated, there
is uncertainty involved in the confidence region's interpretation.

Let $\V{x}_{cm}$ be the operating conditions associated with maximum $g(\V{x},\V{\theta})$ subject to the
constraint that $\V{x}$ is within the experimental region. Denote $\VH{x}_{cm}$ and $\VH{\theta}$ as estimators of $\V{x}_{cm}$ and $\V{\theta}$ respectively, then in practice, $\VH{x}_{cm}$ can be
calculated from $g(\V{x};\VH{\theta})$ using a numerical optimization algorithm, such as the
Nelder-Mead (1965) simplex.
A favorable property of $\V{x}_{cm}$ is its interpretation does not depend on unknown model
parameters, unlike the stationary point or constrained stationary point.
A drawback, however, is that in some instances $\VH{x}_{cm}$ is not a consistent estimator.
Consider the case where $g(\V{x},\V{\theta})$ is continuous and
$\varepsilon_i \overset{iid}{\sim}$ N$(0,\sigma^2)$.
If the model is correct, $\VH{\theta}$ is a least-squares estimator and $\V{x}_{cm}$ is unique,
then $\VH{x}_{cm}$ can be shown to be consistent (Kendall and Stuart, 1979, Chapter 18). To
demonstrate an instance where $\V{x}_{cm}$ is not unique and, therefore, $\VH{x}_{cm}$ is not
consistent, consider the $k=2$
second-order response surface where $\beta_1=\beta_2=0$ and $\beta_{11}=\beta_{22}>0$. If the
experimental region is the two dimensional direct product of the interval $[-a,a]$,
i.e.~$[-a,a]^2$, then the maximum
response within the experimental region occurs at $(-a,-a)$, $(a,-a)$, $(-a,a)$ and $(a,a)$.
A formal assessment of whether $\V{x}_{cm}$ is unique could be made with
tests of the appropriate hypotheses. For example, the likelihood of the symmetric model described
earlier could be investigated by testing $\text{H}_0 \colon \beta_1=\beta_2=0$,
$\beta_{11}=\beta_{22}$. If there is insufficient evidence to reject $\text{H}_0$ further
investigation is warranted.


Construction of a confidence region for $\V{x}_{cm}$ using classical methods would require knowledge
of the sampling distribution of $\VH{x}_{cm}$. As this distribution is unknown it is reasonable to explore
the use of bootstrap confidence region methods that do not require its mathematical derivation.
Unlike the confidence region methods described earlier, the bootstrap approach does not require
normality assumptions on the model errors.

\section{Likelihood-Based Bootstrap Confidence Regions}

Hall (1987, 1992) describes three methods for constructing likelihood-based bootstrap confidence
regions for a $k$-variate parameter vector $\V{\xi}$ given a sample of size $n$, namely, the
percentile-$t$ method, the ordinary percentile (hybrid) method, and the percentile method.
In regards to which of these methods is appropriate when $\V{\xi} \equiv \V{x}_{cm}$, there are
several factors that merit consideration. First, the percentile-$t$ method requires an accurate estimate
for $\V{V}$, the asymptotic covariance matrix of $n^{\frac{1}{2}}(\VH{\xi} - \V{\xi})$, where
$\V{V}$ is assumed to be positive definite. The analytic expression for $\VH{V}$, an estimate of $\V{V}$,
is unknown and accurate estimates using resampling techniques are unlikely to obtain in the small
sample designed experiment setting.
Second, and perhaps more importantly, only the percentile method preserves the range of
$\V{\xi}$ in all cases.  In particularly, for $\V{\xi} \equiv \V{x}_{cm}$, only confidence regions
under the percentile method are guaranteed to be bounded by the experimental region.
Therefore, for the purposes of this report, attention is focused exclusively on the percentile
method to construct likelihood-based bootstrap confidence regions for $\V{x}_{cm}$.


\subsection{Percentile Method Algorithm}
\label{PMAA}
An adaptation of Hall's (1987) likelihood-based bootstrap confidence region algorithm when
$\V{\xi}=\V{x}_{cm}$ is given below, followed by several notes regarding its implementation.

\begin{enumerate}
\item{For the response surface model $\V{y} = \V{X} \V{\theta} + \V{\varepsilon}$,
      determine $\VH{\theta}$ using the method of least-squares.}
\item{Standardize the residual vector $\VH{\varepsilon}$ with the elementwise operation \newline
        $\VH{\varepsilon}_s = \VH{\varepsilon} / \sqrt{\text{diag}(I-X(X'X)^{-1}X')}$.}
\item{Compute $b$ bootstrap estimates of $\V{x}_{cm}$ by following steps 3(a) - 3(d) a total of $b$ times
(see implementation notes for convenient choices of $b$).}
\begin{enumerate}
  \item{Generate a bootstrap sample of $\VH{\varepsilon}_s$, denoted by $\VH{\varepsilon}^*_s$.}
  \item{Calculate the corresponding bootstrap sample of the data with
        $\V{y}^* = \V{X} \VH{\theta} + \VH{\varepsilon}^*_s$.}
  \item{For the response surface model $\V{y}^* = \V{X} \V{\theta}^* + \V{\varepsilon}$,
      determine $\VH{\theta}^*$ using the method of least-squares.}
  \item{Based on $\VH{\theta}^*$, calculate $\VH{x}^*_{cm}$.}
\end{enumerate}
\item{Fit a nonparametric density $\hat{f}$ to the $\VH{x}^*_{cm,i}$, $i=1$,~2,\dots,~$b$.}
\item{The contour on $\hat{f}$ of smallest content that captures
      $100(1-\alpha) \%$ of the $b$ estimates $\VH{x}^*_{cm}$ is a $100(1-\alpha) \%$ likelihood-based
      confidence region for $\V{x}_{cm}$ using the percentile method.}

\end{enumerate}

\subsection{Implementation Notes}
Steps $1-3$ are an application of bootstrapping residuals to generate bootstrap
estimates of $\V{x}_{cm}$. Bootstrapping residuals was proposed by Efron (1979) and is considered
appropriate when $\V{X}$ is fixed and the model is correct with exchangeable errors.
Standardization of the residual vector in step 2 ensures that the variance of each
$\hat{\varepsilon}_{s,i}$ matches that of the unobservable $\varepsilon_i$.
A simple bootstrap sample is generated by drawing a sample of size $n$ with replacement from the
elements of $\VH{\varepsilon}_s$. The balanced bootstrap, proposed by Davison et al. (1986), was used
in the examples and simulation studies of section \ref{simulation} to ensure that
$\overline{\VH{\varepsilon}}^* = \overline{\VH{\varepsilon}}$.
In step 4 the conditional distribution of $\VH{x}_{cm}$ is approximated nonparametrically and used
as the basis for selecting the confidence region boundary.

In practice, the level contour on $\hat{f}$ of smallest content that captures $100(1-\alpha)\%$
of the $b$ estimates $\VH{x}^*_{cm}$ was determined as follows:
Choose $b$ such that $(1-\alpha)b$ is a non-negative integer and assume that
$\hat{f}(\VH{x}_{cm,i}^*)$, $i=1$,~2,\dots,~$b$, are unique. Let $\hat{f}_{\alpha}$ be the value of $\hat{f}$
corresponding to the level contour of smallest content on $\hat{f}(\V{x})$ that captures exactly
$(1-\alpha)b$ of the $\VH{x}^*_{cm,i}$, $i=1$,~2,\dots,~$b$. To calculate $\hat{f}_{\alpha}$ consider
that, by definition, $\hat{f}(\VH{x}_{cm,i}^*) \ge \hat{f}_{\alpha}$ for all $\VH{x}_{cm,i}^*$ captured
by the contour and $\hat{f}(\VH{x}_{cm,i}^*) < \hat{f}_{\alpha}$ for every other $\VH{x}_{cm,i}^*$.
Therefore, $\hat{f}_{\alpha}$ is the $(1-\alpha)b^{th}$ element of the set with elements
$\hat{f}(\VH{x}_{cm,i}^*)$, $i=1$,~2,\dots,~$b$, sorted in descending order. In the event that all
$\hat{f}(\VH{x}_{cm,i}^*)$, $i=1$,~2,\dots,~$b$, are not unique, the contour $\hat{f}_{\alpha}(\V{x})$
may capture more than $(1-\alpha)b$ of the $\VH{x}_{cm,i}^*$, $i=1$,~2,\dots,~$b$, in which case the
confidence region is conservative. After identifying $\hat{f}_{\alpha}$ in this manner the
confidence region boundary was graphically displayed using the GCONTOUR procedure of
$\text{SAS}^{\circledR}$.

\subsection{Application to Second-Order Response Surfaces}
Likelihood-based confidence regions were constructed for $\V{x}_{cm}$, where $g(\V{x},\V{\theta})$ was a
second-order polynomial $(k=2)$ for the following cases:
1) concave down with $\V{x}_{cm}$ located inside the experimental region and
2) saddle system with $\V{x}_{cm}$ located, necessarily, on a boundary of the experimental region.
The experimental region was defined as the two dimensional direct product of the interval $[-1.4,1.4]$.
Model parameters and coordinates of $\V{x}_{cm}$ for both models are summarized in table
\ref{Ta:TrueModels}.

As indicated in section \ref{PMAA}, a nonparametric estimate of the sampling distribution of $\VH{x}_{cm}$
conditional on $b$ bootstrap estimates of $\V{x}_{cm}$ serves as the basis for constructing the
confidence region boundary. A product kernel density estimator for this purpose is given by
\begin{equation} \label{ProductKernel}
\hat{f}(\V{x}) = \frac{1}{bh_1 h_2} \sum_{i=1}^b \left\{ \prod_{j=1}^2
                   K \left( \frac{x_i - \hat{x}_{cm,i,j}^*}
                                 {h_j} \right) \right\} ,
\end{equation}
where $K$ is a univariate kernel and $h_i$ is the bandwidth for the $i$th coordinate
(see, e.g., Silverman, 1986, Scott, 1992). Gasser and M\"uller (1979) showed that if
$K$ is symmetric the bias of $\hat{f}(\V{x})$ can be inflated near the boundaries. M\"uller (1988)
showed that uniform bias over the entire support of $f(\V{x})$ can be achieved if an asymmetric boundary
corrected kernel is used in \eqref{ProductKernel}. Gasser and M\"uller (1979) and Jones (1993), among
others, describe univariate boundary kernels designed for estimating densities with a single boundary.
Densities with support over an interval may be estimated using two such kernels, one at each boundary, but
this is appropriate only if the bandwidth is small relative to the interval length. Hart and Wehrly's
(1992) linear boundary kernel was implemented in \eqref{ProductKernel}, as this kernel is specifically
designed to estimate densities with support on a interval, where the bandwidth need not be small relative
the interval length. The biweight density was used as a basis for constructing this boundary kernel.

Two methods of multivariate bandwidth selection were explored: Normal rule-of-thumb
(see, e.g., Scott, 1992) and the plug-in approach of Wand and Jones (1994, pp 107-113). The sample
standard deviation in each coordinate of $\VH{x}_{cm}^*$ was used to estimate the scale of the
conditional distribution of $\VH{x}_{cm}$, as some measure of scale is required by both bandwidth
selectors. The possibility exists for all bootstrap estimates for $\V{x}_{cm}$ to be located on the
experimental region boundary, a situation that occured almost exclusively with the saddle system. To
ensure nonzero scale estimates, the standard deviation for the $i$th coordinate was calculated from
a modified dataset given by
\begin{equation} \label{jitter}
  z_i =
  \begin{cases}
    \V{x}_{i,cm}^* - \text{sign}(\V{x}_{i,cm}^*) \delta  &\text{if} \quad \text{abs}(\V{x}_{i,cm}^*) = 1.4 \\
    \V{x}_{i,cm}^*  &\text{otherwise} ,
  \end{cases}
\end{equation}
where $\delta$ is a random $U(0,0.05)$ deviate. The effect of using the modified dataset to estimate
standard deviation was negligible in the concave-down system, as most bootstrap estimates did not
fall on a boundary. For the saddle system, using the modified dataset prevented nonzero bandwidth
estimates and, thus, permitted use of the product kernel density estimator.

\subsection{Simulation Study Results}
\label{simulation}
It is of interest to compare the bootstrap confidence region performance under both methods of bandwidth
selection, i.e.~Normal rule-of-thumb and the plug-in approach. Data for 500 experiments were simulated
by adding a $\text{N}(0,3^2)$ deviate to the `true' response at operating conditions associated with
a $k=2$ rotatable central-composite design (5 center runs, $n=13$). Confidence regions using both
bandwidth selectors were constructed using 2000 bootstrap samples in each simulated experiment.
Bandwidths under the plug-in method were slightly larger, on average, than those of the Normal
rule-of-thumb (see table \ref{Ta:Bandwidths}). This result is not unexpected, as simulation studies
of Wand and Jones (1994) indicate a tendency for the multivariate plug-in method to oversmooth
in some cases.

\begin{figure}[p]
\includegraphics{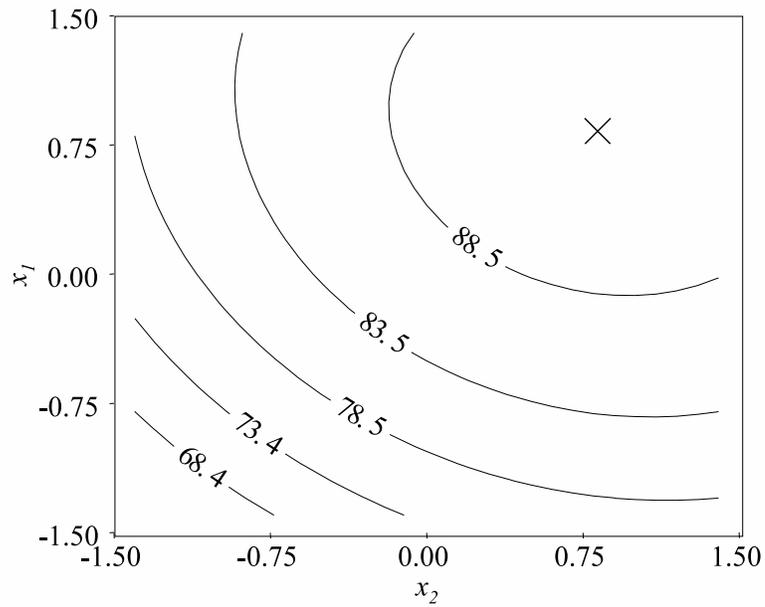}
  \caption{Contour plot of the true concave-down response surface, where $\times$ identifies $\V{x}_{cm}$.}
           \label{Fi:True_Concave_Down_RS}
\end{figure}

\begin{figure}[p]
\includegraphics{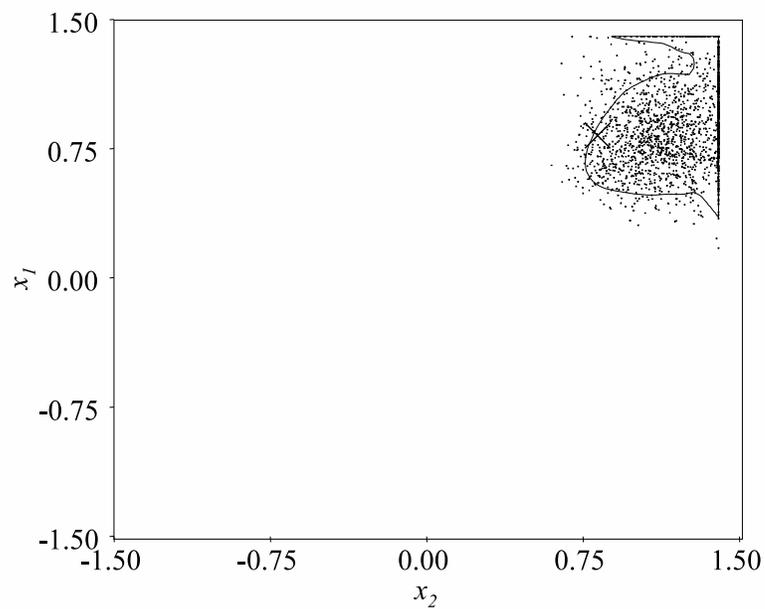}
  \caption{A 90\% confidence region for $\V{x}_{cm}$ where the true response surface is concave-down and
           $\times$ identifies $\V{x}_{cm}$.}
           \label{Fi:Concave_Down_CR}
\end{figure}

A contour of the true concave-down response surface and a representative confidence region for
$\V{x}_{cm}$ with $b=2000$ are provided in figures \ref{Fi:True_Concave_Down_RS} and
\ref{Fi:Concave_Down_CR}, respectively. Bootstrap estimates for $\V{x}_{cm}$ are indicated with
points in figure
\ref{Fi:Concave_Down_CR}, with a solid line identifying the confidence region boundary.
Note that the confidence region boundary is within the experimental region, as required. Had the
hybrid or percentile-$t$ methods been implemented this would not have been true in all cases.

\begin{figure}[p]
\includegraphics{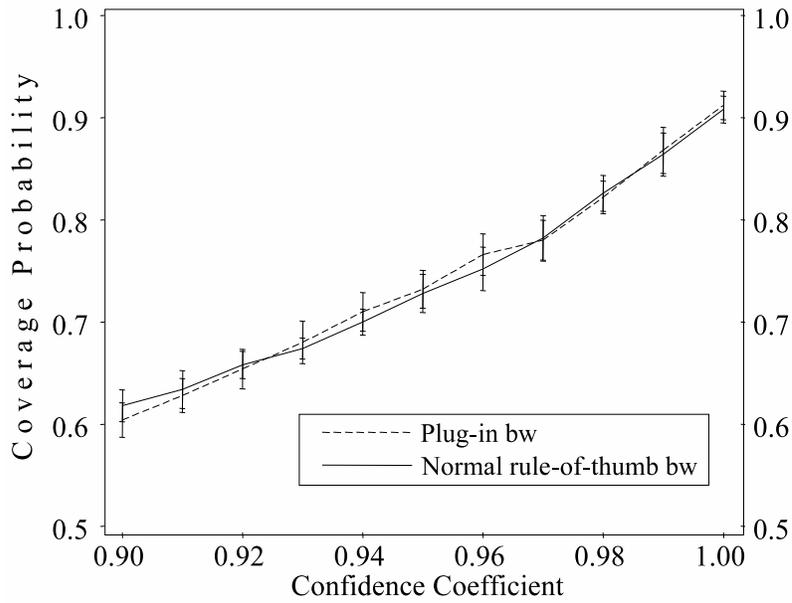}
  \caption{Comparison of coverage probability under the Normal rule-of-thumb and plug-in bandwidth
           selectors. The true response surface is concave-down.}
           \label{Fi:ConcaveDown_BW_comparison}
\end{figure}

The simulated experiments were arranged into five groups of 100 and the coverage probability
calculated in each group for confidence coefficients ranging from $90 - 100\%$. The mean coverage
probability, with error bars placed at one standard error from the mean in either direction,
is plotted in figure \ref{Fi:ConcaveDown_BW_comparison}. Note that
there is little difference in coverage probability under the two approaches, a result expected
from the close similarity in bandwidths seen in table \ref{Ta:Bandwidths}. One approach for
potentially improving the coverage probability is explored in the discussion section.

\begin{figure}[p]
\includegraphics{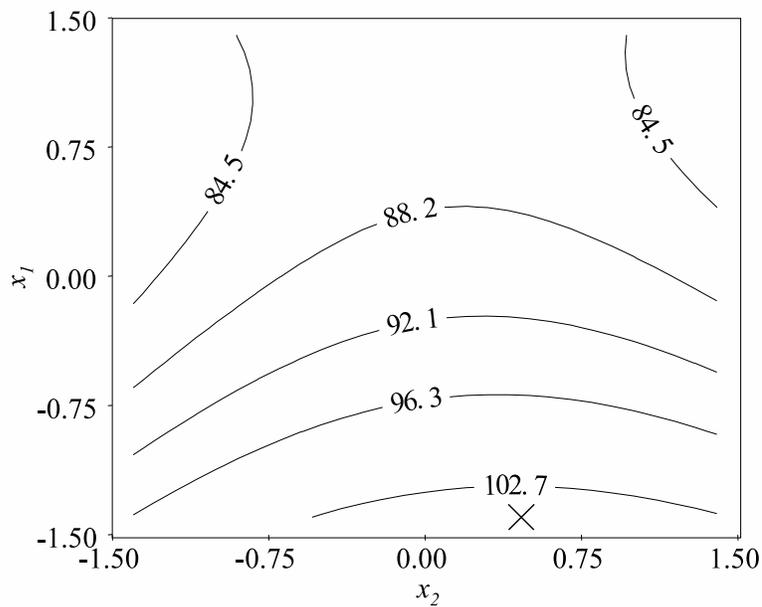}
  \caption{Contour plot of the true saddle system response surface, where $\times$ identifies $\V{x}_{cm}$.}
           \label{Fi:True_Saddle_System_RS}
\end{figure}

\begin{figure}[p]
\includegraphics{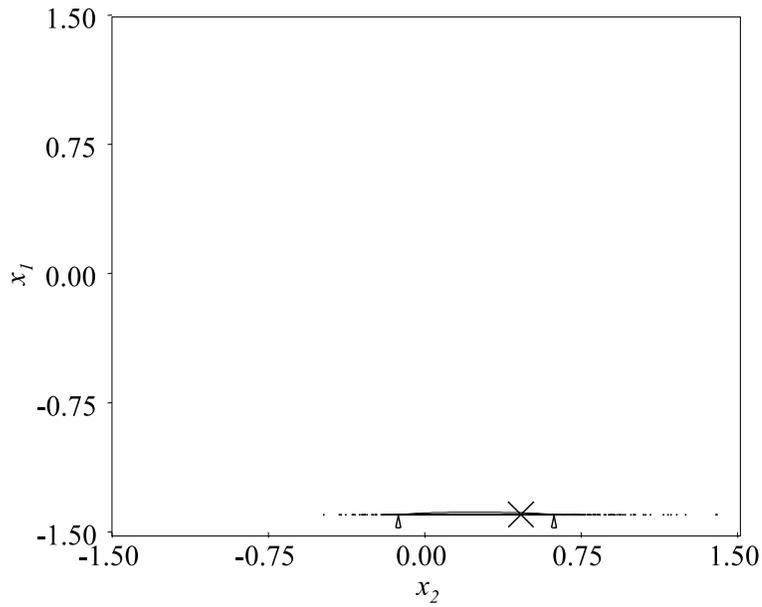}
  \caption{A 90\% confidence region for $\V{x}_{cm}$ where the true response surface is saddle system and
           $\times$ identifies $\V{x}_{cm}$.}
           \label{Fi:Saddle_System_CR}
\end{figure}

A contour plot of the true saddle system and a representative confidence region for $\V{x}_{cm}$ are
provided in figures \ref{Fi:True_Saddle_System_RS} and \ref{Fi:Saddle_System_CR}, respectively.
Note that in this case $\V{x}_{cm}$ is located on the boundary of the experimental region. Since all
bootstrap estimates for $\V{x}_{cm}$ are also on the boundary, visual identification of the confidence
region is difficult. In the figure, two small triangles identify the location of lower and upper
boundaries in the $x_2$ coordinate. In the $x_1$ coordinate the confidence region extends slightly
away from the experimental region boundary.

\begin{figure}[p]
\includegraphics{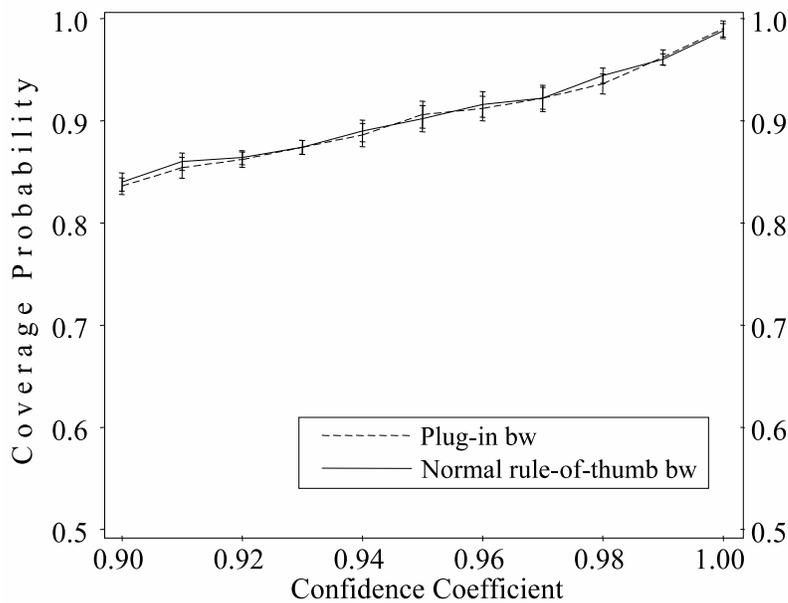}
  \caption{Comparison of coverage probability under the Normal rule-of-thumb and plug-in bandwidth
           selectors. The true response surface is a saddle system.}
           \label{Fi:Saddle_BW_comparison}
\end{figure}

Figure \ref{Fi:Saddle_BW_comparison} summarizes the coverage
probability under both bandwidth selectors for the saddle system. As with the concave-down response
surface, the coverage probability is statistically indistinguishable under the two approaches. The
coverage probability is closer to the nominal level than for the concave-down response surface.

Calculation of the Normal rule-of-thumb bandwidths requires less time and computational resources
than bandwidths under Wand and Jones' plug-in method.
Since the coverage probability was essentially the same under both methods, the
Normal rule-of-thumb method was implemented in the simulation studies that follow.

A simulation study was conducted to assess whether increasing the number of bootstrap samples
would improve the coverage probability. Confidence regions were constructed from the same
simulated data described earlier using $b=4000$ and $b=6000$. In neither the saddle
system nor concave-down case was the coverage probability significantly effected by increasing $b$.

To explore the effect of sample size on coverage probability, confidence regions for $\V{x}_{cm}$
were constructed under both second-order response surfaces for $n=13$, $n=26$ and $n=208$, corresponding
to 1, 2 and 16 experimental replications of a rotatable central-composite design with 5 center runs.
The results for the concave-down and saddle system with $b=2000$ are summarized in
figures \ref{Fi:ConcaveDown_SampleSize_comparison} and \ref{Fi:Saddle_SampleSize_comparison}, respectively.
In both cases the coverage probability approaches the nominal level with increasing sample size.

\section{Discussion}
Existing confidence regions for operating conditions associated with the maximum of a response
surface, either unconstrained or constrained to a specified region, are based on the stationary
point. These approaches are only applicable to second-order models under the assumption of
normally distributed errors. Also, the interpretation of these confidence regions can be ambiguous,
since this requires assessment of the unknown elements of the $\V{B}$ matrix.

In contrast, the interpretation of a likelihood-based bootstrap confidence region for $\V{x}_{cm}$
does not depend on the nature of the response surface, assuming $\V{x}_{cm}$ is unique.
For example, if $g(\V{x},\V{\theta})$ is a second-order polynomial the confidence region
interpretation is the same whether $g(\V{x},\V{\theta})$ is concave-down, concave-up or a saddle
system. In addition, the approach is not restricted to second-order models nor does it require
assumptions on the model error distribution, except for exchangeability of the errors. In principle, the
methodology is also applicable to models where $g(\V{x},\V{\theta})$ is nonlinear in $\V{\theta}$,
though Hjorth (1994, pp 190) indicates that in non-linear regression applications a direct analogue of
standardized residuals is generally not available.

\begin{figure}[t!]
\includegraphics{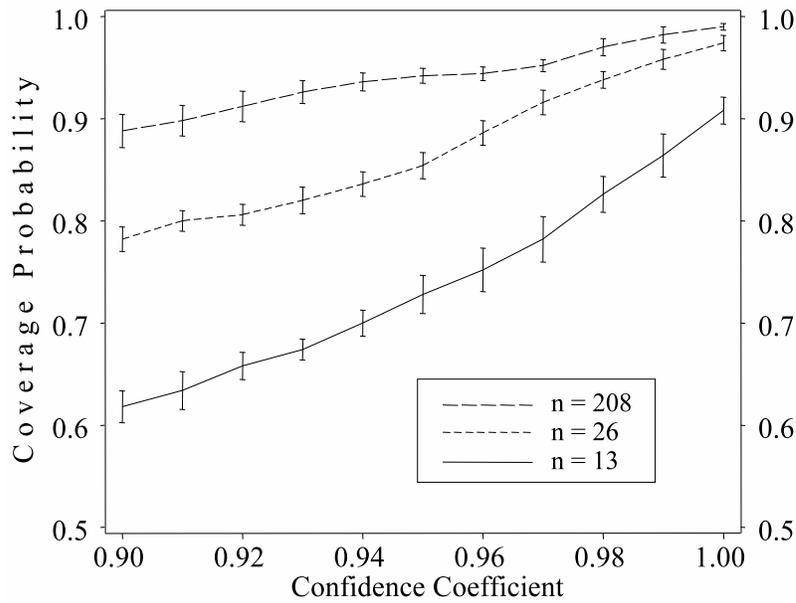}
\vspace*{-6pt}
  \caption{Comparison of coverage probability for three different sample sizes where the response
           surface is concave-down and $b=2000$ with Normal-rule-of-thumb bandwidths.}
           \label{Fi:ConcaveDown_SampleSize_comparison}
\vspace*{-6pt}
\end{figure}

\begin{figure}[b!]
\includegraphics{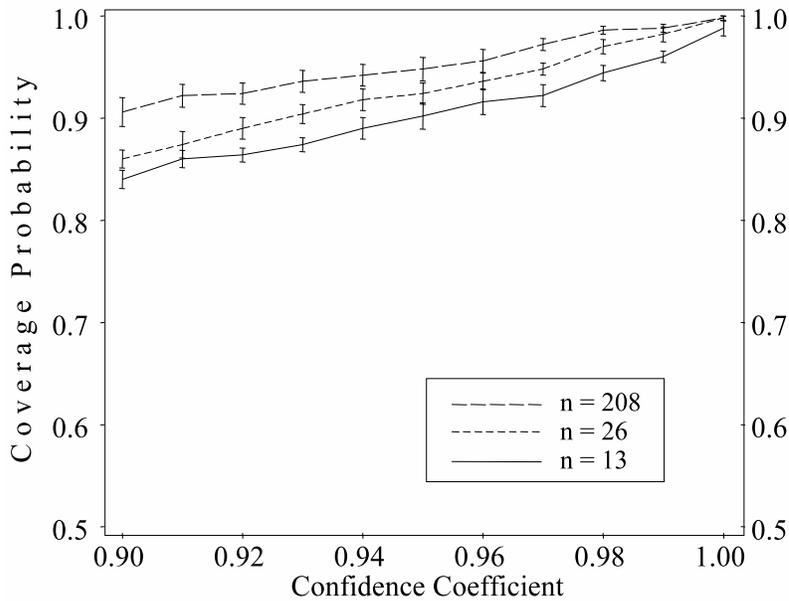}
\vspace*{-6pt}
  \caption{Comparison of coverage probability for three different sample sizes where the response
           surface is a saddle system and $b=2000$ with Normal-rule-of-thumb bandwidths.}
           \label{Fi:Saddle_SampleSize_comparison}
\end{figure}

Simulation results from section \ref{simulation} (see figure \ref{Fi:ConcaveDown_SampleSize_comparison})
indicate the bootstrap confidence region coverage probability
was less than nominal under the concave-down system ($n=13$).
Coverage probability was higher for the saddle system ($n=13$), but still below the nominal level
(see figure \ref{Fi:Saddle_SampleSize_comparison}). A possible reason that higher coverage probability
was observed for the saddle system is the estimated conditional distribution of $\VH{x}_{cm}$
is essentially restricted to one dimension in this case, thereby reducing the
`curse of dimensionality'.
Evidence that the coverage probability in both systems converges to the nominal level with
increasing sample size, an indication of the asymptotic accuracy of the methodology, is apparent
in figures \ref{Fi:ConcaveDown_SampleSize_comparison} and \ref{Fi:Saddle_SampleSize_comparison}.

Loh (1987, 1991) proposed a bootstrap calibration technique to improve the accuracy of confidence
sets. In brief, the approach consists of estimating the confidence coefficient associated with the
desired coverage probability.
For example, to achieve a coverage probability of 90\% in the concave-down response surface
case ($n=13$), it is apparent in figure \ref{Fi:ConcaveDown_SampleSize_comparison} that
a confidence coefficient of approximately 99.5\% is required.
It is also apparent that in this case realization of coverage probabilities greater than
$\approx 91\%$ are not feasible with this approach, unless the sample size is increased.
For the saddle system ($n=13$) a confidence coefficient of approximately 95\% is needed to
achieve a coverage probability of 90\% (see figure \ref{Fi:Saddle_SampleSize_comparison}).

The accuracy of bootstrap confidence regions for $\VH{x}_{cm}$ using the percentile method is
largely determined by how accurately the conditional distribution of $\VH{x}_{cm}$, $f$,
is estimated. Under the kernel method implemented in section 2, choice of bandwidth has a critical
impact on this accuracy. Hall (1987) indicated that bandwidths which are optimal in a global sense,
such as minimization of the mean integrated squared error of $\hat{f}$, can produce confidence
regions that are ``unduly bumpy''. He noted that this can be attributed to the fact that the
ratio of the variance to the squared bias of $\hat{f}$ is greater in the tails than in the
central region of the distribution. Therefore, the Normal rule-of-thumb and plug-in bandwidth
selectors were investigated for their tendency to avoid undersmoothing. The $k=2$ concave-down
and saddle system cases explored in section \ref{simulation} are examples where these methods provide
reasonable bandwidth estimates. Jones (1993) followed a similar approach when he used a plug-in
bandwidth selector in conjunction with a univariate boundary kernel.

There are situations, however, in which bandwidth selection is more challenging. For example,
if $f$ is bimodal, use of the standard deviation to estimate the scale of $f$ can result in
considerable oversmoothing. Janssen et al. (1995) developed more robust scale estimators for
such cases. Consider, also, a situation where all bootstrap estimates for $\V{x}_{cm}$ are equally
distributed on the two boundaries $\mathcal{A} = \{\V{x}\colon x_1=1.4, x_2 \in (-1.4, 1.4) \}$
and $\mathcal{B} = \{\V{x} \colon x_1 \in (-1.4,1.4), x_2 = 1.4 \}$.
On boundary $\mathcal{A}$ bandwidth $h_1$ should be much smaller than $h_2$.
However, on boundary $\mathcal{B}$ the opposite is true.
In this case, a variable kernel density estimator would seem more appropriate than the fixed
bandwidth kernel estimator implemented in section \ref{simulation}, as this would allow for different levels of smoothing depending on the location of bootstrap estimates for $\V{x}_{cm}$.

Application of the bootstrap confidence region methodology was restricted to the $k=2$ case
where the experimental region was rectangular in shape. The bootstrap and kernel density estimation
techniques described in section 2 are not limited to two dimensions. Wand and Jones'
plug-in bandwidth selector can also be extended to higher dimensions, though with greater
computational expense. In light of the close comparison of coverage probability
under the Normal rule-of-thumb and plug-in methods observed in section \ref{simulation}, rule-of-thumb
methods may be adequate in many higher-dimensional applications. Graphical presentation of
confidence regions for $k>3$ is a challenging issue, though this is not unique to our application.
Staniswalis, Messer and Finston (1993) describe a multivariate boundary kernel designed for
regions of arbitrary shape which obtains the correct order of bias over the entire region
(see, also, Scott, 1992, pp 155).

Throughout this article discussion has focused on maximizing a single response.
However, researchers in many areas of application are faced with the problem of simultaneously
improving multiple responses that depend on a common set of controllable variables.
Since a single operating condition is rarely optimal for all responses, compromise must be
incorporated into the estimation procedure. Desirability optimization methodology
(Harrington, 1965, Derringer and Suich, 1980) addresses this problem and has been proven
effective in a wide range of applications involving continuous responses. In brief, the
approach consists of estimating the operating conditions that maximize $D$, a simultaneous
measure of the desirability of all the responses. An arguable shortcoming of the methodology
is it does not provide for estimation of the variability of these operating conditions.
Bootstrap confidence region methodology is defined in sufficiently general terms to
allow for construction of a confidence region for the operating condition
that maximizes $D$ constrained to the experimental region, under the assumption that this
parameter is unique.

Of practical interest is the computational resources and time required to construct a likelihood-based
bootstrap confidence region. The programs that generated the results of section \ref{simulation}
were written in $\text{SAS}^{\circledR}$ and run on a 300 megahertz $\text{Pentium}^{\circledR}$
II desktop computer with 128 megabytes of RAM. Approximately 2 minutes is required to
generate a single confidence region with $b=2000$, of which about 25 seconds
are used for steps $1-3$ with the balance of the time taken in steps 4 and 5.
Wand (1994) describes methods of accelerating computations required for multivariate bandwidth
optimization and kernel density estimation. These methods were not implemented
due to technical limitations of $\text{SAS}^{\circledR}$. It is expected that steps 4 and 5
would require considerably less time under their optimized procedures.

\vfill\pagebreak
\begin{center}
\textbf{\large{References}}
\end{center}

\setlength{\hangindent}{10pt} \noindent
Box, G.E.P., and Hunter, J.S. (1954), ``A Confidence Region for the Solution of a Set of
Simultaneous Equations With An Application to Experimental Design,''
\emph{Biometrika}, 41, 190-199.

\setlength{\hangindent}{10pt} \noindent
Davison, A.C., Hinkley, D.V., and Schechtman, E. (1986), ``Efficient Bootstrap Simulation,''
\emph{Biometrika}, 73, 555-566.

\setlength{\hangindent}{10pt} \noindent
Derringer, G., and Suich, R. (1980), ``Simultaneous Optimization of Several Response Variables,''
\emph{Journal of Quality Technology}, 12, 214-219.

\setlength{\hangindent}{10pt} \noindent
Draper, N.R. (1963), ``Ridge Analysis for Response Surfaces,''
\emph{Technometrics}, 5, 469-479.

\setlength{\hangindent}{10pt} \noindent
Efron, B. (1979), ``Bootstrap Methods: Another Look at the Jackknife,''
\emph{The Annals of Statistics}, 7, 1-26.

\setlength{\hangindent}{10pt} \noindent
Gasser, T., and M\"uller, H.-G (1979), ``Kernel Estimation of Regression Functions,'' In
\emph{Smoothing Techniques for Curve Estimation} (eds. T. Gasser and M. Rosenblatt),
Heidelberg: Springer-Verlag, pp 23-69.

\setlength{\hangindent}{10pt} \noindent
Hall, P. (1987), ``On the Bootstrap and Likelihood-Based Confidence Regions,''
\emph{Biometrika}, 74, 481-493.

\setlength{\hangindent}{10pt} \noindent
Hall, P. (1992), \emph{The Bootstrap and Edgeworth Expansion},
New York: Springer-Verlag.

\setlength{\hangindent}{10pt} \noindent
Harrington, E.C. (1965), ``The Desirability Function,''
\emph{Industrial Quality Control}, 21, 494-498.

\setlength{\hangindent}{10pt} \noindent
Hart, J.D. and Wehrly, T.E. (1992), ``Kernel Regression When the Boundary Region is Large, With An
Application to Testing the Adequacy of Polynomial Models,''
\emph{The Journal of the American Statistical Assocication}, 87, 1018-1024.

\setlength{\hangindent}{10pt} \noindent
Hoerl, A.E. (1959), ``Optimum Solution of Many Variables Equations,''
\emph{Chemical Engineering Progress}, 55, 67-78.

\setlength{\hangindent}{10pt} \noindent
Hjorth, J.S.U. (1994),
\emph{Computer Intensive Statistical Methods, Validation Model Selection and Bootstrap},
London: Chapman \& Hall.

\setlength{\hangindent}{10pt} \noindent
Janssen, P., Marron, J.S., Veraverbeke, N., and Sarle, W. (1995), ``Scale Measures for Bandwidth Selection,''
\emph{The Journal of Nonparametric Statistics}, 5, 359-380.

\setlength{\hangindent}{10pt} \noindent
Jones, M.C. (1993), ``Simple Boundary Correction for Kernel Density Estimation,''
\emph{Statistics and Computing}, 3, 135-146.

\setlength{\hangindent}{10pt} \noindent
Kendall, M. and Stuart, A. (1979),
\emph{The Advanced Theory of Statistics, Volume II: Inference and Relationship,}
4th edition, New York: Macmillan.

\setlength{\hangindent}{10pt} \noindent
Loh, W.-Y. (1987), ``Calibrating Confidence Coefficients,''
\emph{The Journal of the American Statistical Assocication}, 82, 155-162.

\setlength{\hangindent}{10pt} \noindent
Loh, W.-Y. (1991), ``Bootstrap Calibration for Confidence Interval Construction and Selection,''
\emph{Statistica Sinica}, 1, 479-495.

\setlength{\hangindent}{10pt} \noindent
M\"uller, H.G. (1988).
\emph{Lecture Notes in Mathematics: Nonparametric Regression Analysis of Longitudinal Data.}
New York: Springer-Verlag, 46.

\setlength{\hangindent}{10pt} \noindent
Nelder, J.A., and Mead, R. (1965), ``A Simplex Method for Function Minimization,''
\emph{Computer Journal}, 7, 308-313.

\setlength{\hangindent}{10pt} \noindent
Peterson, J.J. (1992), ``Confidence Regions for Constrained Response Surface Optima,''
Presented at the August, 1992 Joint Statistical Meetings, Atlanta, GA.

\setlength{\hangindent}{10pt} \noindent
Scott, D.W. (1992), \emph{Multivariate Density Estimation: Theory, Practice and Visualization},
New York: Wiley.

\setlength{\hangindent}{10pt} \noindent
Silverman, B.W. (1986), \emph{Density Estimation for Statistics and Data Analysis},
London: Chapman \& Hall.

\setlength{\hangindent}{10pt} \noindent
Stablein, D.M., Carter, W.H., and Wampler, G.L. (1983),
``Confidence Regions for Constrained Optima in Response-Surface Experiments,''
\emph{Biometrics}, 39, 759-763.

\setlength{\hangindent}{10pt} \noindent
Staniswalis, J.G., Messer, K., and Finston, D.R. (1993), ``Kernel Estimators for Multivariate
Regression,'' \emph{Journal of Nonparametric Statistics} \textbf{3}, 103-121.

\setlength{\hangindent}{10pt} \noindent
Wand, M.P. (1994), ``Fast Computation of Multivariate Kernel Estimators,''
\emph{Journal of Computational and Graphical Statistics}, 3, 433-445.

\setlength{\hangindent}{10pt} \noindent
Wand, M.P., and Jones M.C. (1994), ``Multivariate Plug-In Bandwidth Selection,''
\emph{Computational Statistics Quarterly}, 9, 97-116.

\begin{table}[h]
\caption{True second-order model parameters and associated $\V{x}_{cm}$.}
 \begin{center}
   \begin{tabular}{l c c c c c c c} \hline
   Response & & & & & & & \\ \cline{2-7}
   Surface & $\beta_0$ & $\beta_1$ & $\beta_2$ & $\beta_{12}$ & $\beta_{11}$ & $\beta_{22}$ & $\V{x}_{cm}$ \\ \hline
   concave down     & 86.850 & ~5.242 & 4.778 & -0.775 & -2.781 & -2.524 & $(0.828,0.819)$ \\
   saddle           & 90.259 & -6.425 & 1.244 & -0.775 & ~2.781 & -2.524 & $(-1.4,0.462)$ \\ \hline
   \end{tabular}
 \end{center}
 \label{Ta:TrueModels}
\end{table}

\vspace{0.25in}

\begin{table}[h!]
 \caption{Mean and standard error (in parenthesis) of bandwidths estimated from 500 simulated
          response surfaces using the Normal rule-of-thumb and plug-in methods.}
 \begin{center}
 \begin{tabular}{l c c c c c} \hline
 & \multicolumn{2}{c}{Concave-down} &  & \multicolumn{2}{c}{Saddle system} \\ \cline  {2-3} \cline{5-6} Bandwidth selector   & $h_1$ & $h_2$ &  & $h_1$ & $h_2$ \\ \hline
   Normal rule-of-thumb & 0.196   &  0.213   &  & 0.011                 & 0.261   \\
                        &(0.0033) & (0.0037) &  &($6.9 \times 10^{-6}$) &(0.0049) \\ \hline
   Plug-in method       & 0.214   &  0.233   &  & 0.013                 & 0.313   \\
                        &(0.0034) & (0.0038) &  &($13 \times 10^{-6}$)  &(0.0060) \\ \hline
   \end{tabular}
 \end{center}
 \label{Ta:Bandwidths}
\end{table}

\end{document}